\documentstyle[prb,aps,epsf,floats,twocolumn]{revtex}

\begin{document}
\draft

%%%%%%%%%%%%%%%%%%%%%
\author{Akihisa Koga and Norio Kawakami}
%%%%%%%%%%%%%%%%%%%%
\address{Department of Applied Physics, 
Osaka University, Suita, Osaka 565-0871, Japan}
\title{
Excitation Spectrum for Quantum  Spin Systems 
with Ladder, Plaquette and Mixed-Spin Structures}
%%%%%%%%%%%%%%%%%%%%%%%%%%%%%%%%%%%%%%%

\date{\today}
\sloppy
\maketitle

\begin{abstract}
By using the series expansion techniques, we study 
the excitation spectrum for the two-dimensional quantum spin systems 
with ladder, plaquette and mixed-spin structures.
We calculate the spin excitation gap and thus 
determine the phase boundary between
the spin-gap phase  and  the magnetically ordered phase 
rather precisely. 
It is found that the 
phase diagram obtained improves fairly well
the one previously obtained via the ground-state 
susceptibility.
\end{abstract}

\narrowtext
%%%%%%%%%%%%%%%%%%%%%%%%%%%%%%%%%%%%%
\section{Introduction}
%%%%%%%%%%%%%%%%%%%%%%%%%%%%%%%%%%%%%
Two-dimensional (2D) antiferromagnetic 
quantum spin systems with spin gap provide a 
new interesting paradigm for quantum phase transitions.
A typical example is the spin plaquette system such as $\rm CaV_4O_9$,
\cite{CaV4O91,meta,CaV4O92,Troyer,KatoImadapla,GelCVO,CaV4O93,Fukumoto,metagel}
which may be described by the 2D Heisenberg model 
with a plaquette structure.
Another interesting example found recently 
is the 2D spin system composed of 
orthogonal spin dimers such as $\rm SrCu_2(BO_3)_2$,
\cite{Kageyama,Miyahara,BO3}
which may be described by the 2D Heisenberg model 
on a square lattice with some diagonal exchange couplings. 
In these 2D spin systems, the plaquette or dimer structure 
is essential to stabilize the non-magnetic phase with the spin gap.
Concerning the spin gap formation, the topological nature of 
spins is also important in low-dimensional systems. In this context, 
mixed-spin systems have attracted considerable
attention recently, in which the spatial arrangement
of different spins plays a crucial role to generate the spin gap
or induce an antiferromagnetic long-range order.
For instance, see the references 
for experiments\cite{Yee} and 
theories\cite{Pati,Vega,Tonegawa,Fukui,first} in 1D cases.

In the previous paper,\cite{previous} 
we have investigated the ground state quantities 
for the 2D spin systems with ladder, plaquette and mixed-spin structures,
by extending the works\cite{GelCVO,Fukumoto,metagel,iso_ladder} 
based on the series expansion methods.\cite{singh}
Although the quantum phase transitions have been described qualitatively well,
it has turned out that the obtained results lead to
unsatisfactory estimates for the phase boundary 
in some region of the phase diagram.\cite{previous} 
A more crucial problem raised is to what extent
our series expansion correctly captures the lattice 
structure and/or the spin topological properties,
since our expansion approach has relied on the lower-order 
expansions in coupling constants.
Not only to resolve this problem but also  to confirm our approach to be 
reliable, it is desirable to produce more accurate results
by improving the series expansions, and also to
study other quantities besides the ground state quantities.

The purpose of this work is to study the excitation spectrum for
the 2D quantum spin systems with the above-mentioned structures,
 and clarify  the role of 
the competing interactions in the disordered 
phase with the spin gap. We shall see that the excitation 
spectrum calculated in higher orders than the ground-state
susceptibility improves the phase diagram, and
at the same time confirms that our series-expansion approach 
indeed provides reliable results. 

The paper is organized as follows. 
In \S 2, we briefly summarize how to apply 
the series expansion techniques\cite{studychi} to our systems.
By performing the series expansion for the excited states\cite{studygap} and 
employing the asymptotic analysis of power-series expansions\cite{Pade} 
in \S 3, we obtain the dispersion relations and the phase diagram.
We then discuss how the spin gap phase competes with 
the magnetically ordered phase for three kinds of 2D quantum 
spin systems mentioned above.
The last section is devoted to brief summary.

%%%%%%%%%%%%%%%%%%%%%%%%%%%%%%%%%%%%%
\section{Series Expansion Methods}
%%%%%%%%%%%%%%%%%%%%%%%%%%%%%%%%%%%%%

We begin by briefly summarizing the series expansion 
method.\cite{studychi,studygap}
We employ here the cluster expansion
around a given strong-coupling spin singlet state.
Let us explain the idea taking the dimer expansion as an
example.\cite{BO3,previous,iso_ladder,singh,studychi,2D,bilayer} 
First the 2D Hamiltonian is divided into two parts: 
$H=H_0 + H_1$.
The unperturbed Hamiltonian $H_0$ is composed of an assembly of 
isolated singlet dimers which are formed by the strong
antiferromagnetic bonds. Namely, our starting configuration  for the
perturbation has  the  disordered ground state with the spin gap.
We then introduce the interaction term $H_1$ 
among the independent dimers and observe  how the 
physical quantities are changed by exploiting the power 
series expansion with respect to $H_1$.
The advantage of this cluster expansion is that we can 
combine analytical and numerical techniques in 
a complementary way. For example, the computer can be  utilized
to systematically generate the higher order terms from 
the lower ones.\cite{singh,studychi}  We exploit  
the cluster expansions which may be most appropriate for each system
with different structures.

To discuss how the introduction 
of $H_1$ perturbs the disordered state with 
the spin gap and enhances the antiferromagnetic correlation,
we calculate the dispersion relation $E({\bf k})$ for 2D
spin systems with various structures.\cite{previous}
This quantity is expanded as a 
power series in $\lambda$ as
%%%%%%%%%%%%%%%%%%%%%%%%%%%%%%%%%%%%
\begin{eqnarray}
E({\bf k}) &=& \sum_{l,m,n}a_{lmn}\cos (l k_x+ m k_y) \lambda^n,
\end{eqnarray}
%%%%%%%%%%%%%%%%%%%%%%%%%%%%%%%
where the wave number is denoted by ${\bf k} =(k_x, k_y)$ and 
the Brillouin zone for each model will be 
 defined in the following section. 
We shall calculate the dispersion relation up to 
the eighth order in $\lambda$ 
for the ladder-structure system and the fifth order 
for both the plaquette system and the mixed-spin system.
To estimate the minimum value of the dispersion relation in the first 
Brillouin zone, we can also expand the spin gap $\Delta$ up to the 
same order. Since we are not able to analyze 
the critical phenomena only with the obtained power-series,
further asymptotic analyses are necessary to discuss the 
phase transitions. To this end, 
we make use of the Pad\'e approximants and the 
differential methods\cite{Pade} 
to estimate the critical point for the phase transition,
the dispersion relation, etc.
Especially, the critical point between the magnetically ordered and 
disordered phase is estimated not only by the ordinary Dlog 
Pad\'e approximants  but also by the biased Pad\'e approximants. 
In the biased method we assume that the 
phase transition in our 2D quantum spin model 
should belong to the universality class of the 3D classical 
Heisenberg model.\cite{CHN}
Namely,  the critical value $\lambda_c$ for the
perturbation parameter is determined by the formula 
$\Delta\sim (\lambda_c-\lambda)^\nu$ with 
the known exponent $\nu=0.71$ around the transition  point.\cite{Ferer}
We also apply the first-order inhomogeneous differential method 
to the power-series to obtain the dispersion relation. It 
should be noted here that 
since higher-order coefficients in the series expansions 
are necessary to deduce
the dispersion relation in this method correctly,
we might be  sometimes  left with  wrong values at a certain wave number
 after applying the asymptotic analysis.
It is known that this type of pathology occasionally happens
in these asymptotic approximations.\cite{Pade}
If we  carefully 
discard this spurious behavior to find the correct one, 
these analyses provide a fairly good approximation in many cases,
which will be explicitly shown in each case treated below.

%%%%%%%%%%%%%%%%%%%%%%%%%%%%%%%%%%%%%%%%%%%%%%%%%%%%%%%%%%%%%%%%%%%%%%%%
\section{Excitation Spectrum and Phase Diagram}
%%%%%%%%%%%%%%%%%%%%%%%%%%%%%%%%%%%%%%%%%%%%%%%%%%%%%%%%%%%%%%%%%%%%%%%%

Let us now introduce the 2D antiferromagnetic quantum spin system 
defined by the Heisenberg Hamiltonian, 
%%%%%%%%%%%%%%%%%%%%%%%%%%%%%%%%
\begin{eqnarray}
H & =  & H_0 + H_1, \\
H_0 & = & J_1 \sum_{(i,j)\in G_1}{\bf S}_{i} \cdot {\bf S}_{j}, \\
H_1 & = & J_2 \sum_{(i,j)\in G_2} {\bf S}_{i} \cdot {\bf S}_{j} +
J_3 \sum_{(i,j)\in G_3} {\bf S}_{i} \cdot {\bf S}_{j},
\end{eqnarray}
%%%%%%%%%%%%%%%%%%%%%%%%%%
where  $J_{1}$, $J_{2}$ and $J_{3}$ denote 
the antiferromagnetic coupling constants, 
and  ${\bf S}_j$ is   the spin operator at the $j$-th site. 
To treat the mixed-spin systems as well as 
the ladder and plaquette systems,  the spin ${\bf S}_j$ 
is allowed to take different values at each cite.
We denote  the  bonds $(i,j)$ for the non-perturbed Hamiltonian as 
$G_1$, while those for the perturbed parts  as $G_2$ and $G_3$.
By appropriately choosing the set of ($G_1$, $G_2$, $G_3$),
we can deal with the 2D systems with various structures by
the series expansion techniques.
We  treat below the case of $\lambda (\equiv J_2/J_1)<1$ and 
$\alpha\lambda (\equiv J_3/J_1)<1 (0<\alpha<1)$.
In the following, the excitation spectrum is analyzed to
discuss the quantum phase transitions 
for the 2D antiferromagnetic spin systems with ladder, plaquette 
and mixed-spin structures.  We carry out the 
dimer expansion, the plaquette expansion and
the mixed-spin cluster expansion.
Starting with the above strong-coupling spin singlet states, 
we can perform the cluster expansion with respect to $\lambda$ 
and $\alpha\lambda$. 

\subsection{Dimer expansion for ladder-structure systems}

\begin{figure}[htb]
\vspace{0.1cm}
\epsfxsize=7cm
\centerline{\epsfbox{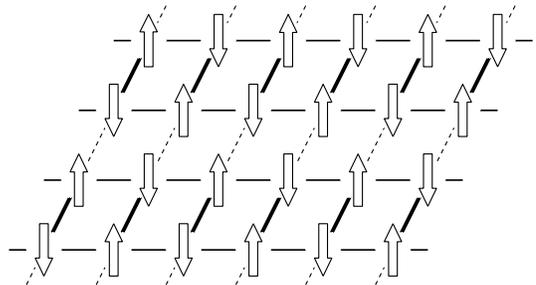}} 
\caption{ The  2D $s=1/2$ spin system with ladder structure.
See the text as for    
the meaning of  the bold, the thin, and the dashed lines.
}
\label{fig:model_d}%-----------------------------------------
\end{figure}

\begin{figure}[htb]
\vspace{0.1cm}
\epsfxsize=7cm
\centerline{\epsfbox{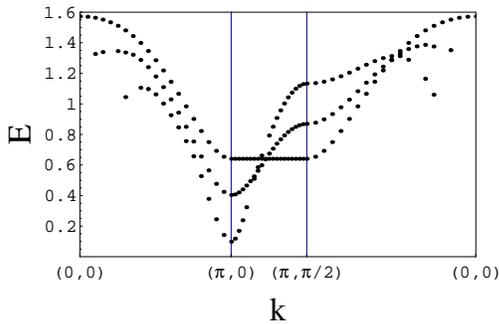}} 
\caption{Plots of the spin-triplet excitation spectrum $E({\bf k})$
along high-symmetry cuts through the Brillouin zone 
for the system with the couplings $\alpha_{\rm L}=0.0, 0.5, 1.0$
(shown in the figure from the top to the bottom at $(\pi, 0)$, 
respectively) when $\lambda_{\rm L}=0.5$.
}
\label{fig:dis_d}%-----------------------------------------
\end{figure}

\begin{figure}[htb]
\vspace{0.1cm}
\epsfxsize=7cm
\centerline{\epsfbox{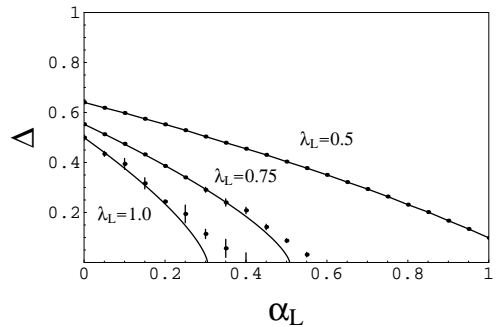}} 
\caption{Spin gap $\Delta=E(\pi, 0)$ for the coupled-ladder system 
in Fig. \protect{\ref{fig:model_d}}. 
The dots with error bars are the 
results obtained by the first-order differential method, whereas the
solid lines denote the corrected values by applying the 
Pad\'e approximants to the spin gap around the critical point
complementarily.}
\label{fig:gap_d}%-----------------------------------------
\end{figure}

%%%%%%%%%%%%%%%%%%%%%%%%%%%%%%%%%%%%
We first discuss a 2D spin system with the ladder structure, 
which is shown schematically in Fig. \ref{fig:model_d}, where 
the bold, the thin and the dashed lines represent the coupling 
constants $1$, $\lambda_{\rm L}$ and 
$\alpha_{\rm L} \lambda_{\rm L} $, respectively.
%%%%%%%%%%%%%%%%%%%
% --------- figure model for dimer expansion ----------------
%%%%%%%%%%%%%%%%%
It is noted here that the spin ladder system ($\alpha_{\rm L}=0$) 
was already studied
in detail by the cluster expansion.\cite{iso_ladder}
By changing $\alpha_{\rm L}$, 
we can see how  the isolated 2-leg ladder $(\alpha_{\rm L}=0)$ 
is changed to  the 2D system. 
We calculate the energy for spin-triplet excitations
 by means of the dimer expansion 
up to the eighth order in $\lambda_{\rm L}$ 
for various values of $\alpha_{\rm L}$. Note that 
the Brillouin zone is reduced to half of the original one
because  the dimer singlet is composed of two spins in the $y$-direction.
By applying the first-order inhomogeneous differential method\cite{Pade}
to the  power series computed above,
we obtain the spin-triplet excitation spectrum  
shown in Fig. \ref{fig:dis_d} in the case of $\lambda_{\rm L}=0.5$.
%%%%%%%%%%%%%%%%%%%%%%%%%%%%%%%%%
% --------- dispersion relation for dimer expansion ----------------
%%%%%%%%%%%%%%%%%%%%%%%%%%%%%%%%%
When $\alpha_{\rm L}=0$, the system is reduced to the isolated 2-leg ladders 
with the inter-leg (intra-leg) coupling constant $1 (\lambda_{\rm L}=0.5)$,
which is known to have the disordered ground state with the spin gap.
\cite{Rice}
This gives rise to the flat dispersion between $(\pi,0)$ and $(\pi,\pi/2)$.
The computed coefficients for the spin gap $\Delta= E(\pi, 0)$,
in the series of $\lambda_{\rm L}$ are tabulated 
for some particular values of $\alpha_{\rm L}$ in Table \ref{tbl:I}.
Note that  for the isolated ladder case $(\alpha_{\rm L}=0)$, 
our results correspond to those obtained previously by 
Oitmaa {\it et al.}\cite{iso_ladder}.
%%%%%%%%%%%%%%%%%%%%%%%%%%%%%%%%%%%%%%%%%%%%%%%%%%%%
\begin{table}
\caption{Series coefficients for the dimer expansion of 
the spin gap $\Delta = E(\pi, 0)$ 
for the coupled-ladder system.}
\begin{tabular}{ccccc}
n & $\alpha_{\rm L}=0.0$ & $\alpha_{\rm L}=0.2$ & 
$\alpha_{\rm L}=0.5$ & $\alpha_{\rm L}=1.0$ \\
\hline
0 & 1.0000000  & 1.0000000 & 1.0000000 & 1.0000000\\
1 &-1.0000000  &-1.1000000 &-1.2500000 &-1.5000000\\
2 & 0.50000000 & 0.38500000& 0.15625000&-0.37500000\\
3 & 0.25000000 & 0.16025000&0.066406250& 0.031250000\\
4 &-0.12500000 &-0.17002083&-0.22737630&-0.34635417\\
5 &-0.27343750 &-0.22812954&-0.26126692&-0.88050673\\
6 &-0.15332031&-0.041639568&0.018219038&-0.16209751\\
7 & 0.24560547 & 0.30931367& 0.36472170&-0.20085681\\
8 & 0.48133850 & 0.37176730& 0.34136063&-0.63949210\\
\end{tabular}
\label{tbl:I}
\end{table}
%%%%%%%%%%%%%%%%%%%%%%%%%%%%%%%%%%%%%%%%%%%%%%%%%%%%
The obtained  spin gap with a fixed $\lambda_{\rm L}$ is shown 
in Fig. \ref{fig:gap_d}.
%%%%%%%%%%%%%%%%%%%%%%%%%%%%%%%%%
% --------- gap for dimer expansion ----------------
%%%%%%%%%%%%%%%%%%%%%%%%%%%%%%%%%
It is seen that the spin gap  decreases with the increase of 
the inter-ladder coupling $\alpha_{\rm L}$  and finally 
vanishes at which the quantum phase transition 
to the antiferromagnetically ordered state occurs.
We wish to mention  that the order in  our cluster expansion is 
not high enough to deduce the accurate dispersion for $\alpha_{\rm L}$
close to the transition point within the first-order inhomogeneous 
differential approximation,  as seen in Fig. \ref{fig:gap_d}.  It thus
seems difficult to deduce the critical point $\alpha_c$ correctly.
However, as far as the critical value is concerned, we can use alternative
analysis based on the Pad\'e approximants, which provides a rather  
accurate estimate for $\alpha_c$,  by assuming 
$\Delta \sim (\alpha_{c}-\alpha)^{\nu}$ near the critical point.
By employing the latter  analysis complementarily around the
critical point  $\alpha_c$, we have obtained 
the corrected  spin gap as a function of 
the inter-ladder coupling  $\alpha_{\rm L}$, which is
shown as the solid line in Fig. \ref{fig:gap_d}.

\begin{figure}[htb]
\vspace{0.1cm}
\epsfxsize=7cm
\centerline{\epsfbox{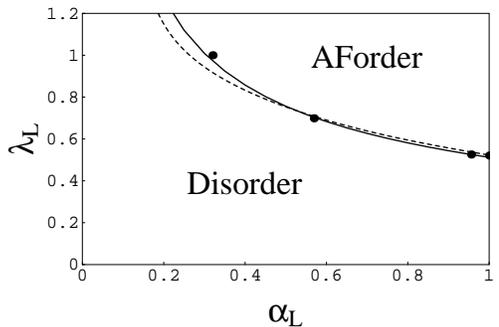}} 
\caption{Phase diagram for the coupled-ladder system in Fig. 
\protect{\ref{fig:model_d}}.
The solid line indicates the phase boundary obtained 
by biased [4/3] Pad\'e approximants for the spin gap while
the dashed line have been obtained in our previous paper.
\protect{\cite{previous}}
The solid circles represent the results of 
the QMC simulations.
\protect{\cite{KatoImada,ImadaIino,Nonomura}}
}
\label{fig:phase_d}%-----------------------------------------
\end{figure}

We also show the phase diagram for the coupled-ladder system
 in Fig. \ref{fig:phase_d}. 
%%%%%%%%%%%%%%%%%%%%%%%%%%%%%%%%%
% --------- phase diagram for dimer expansion ----------------
%%%%%%%%%%%%%%%%%%%%%%%%%%%%%%%%%
The solid line represents the phase boundary obtained by 
the biased [4/3] Pad\'e approximants for the spin gap, and
the dashed line is the
boundary determined previously by the staggered susceptibility\cite{previous}.
  It is remarkable that 
the present result is in fairly good agreement with the QMC simulations,
\cite{KatoImada,ImadaIino,Nonomura} and considerably 
improves the previous one 
especially in the region with small $\alpha_{\rm L}$.

\begin{figure}[htb]
\vspace{0.1cm}
\epsfxsize=7cm
\centerline{\epsfbox{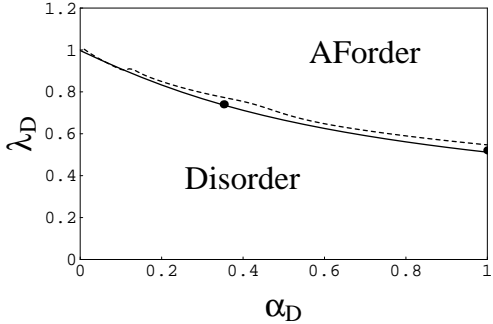}} 
\caption{Phase diagram for the coupled-dimer-chain system  
in Fig. \protect{\ref{fig:model_d}}.  
The solid line indicates the 
phase boundary obtained by the biased [4/3] Pad\'e approximants.
We also show the phase boundary which was determined by applying 
Dlog [4/3] Pad\'e approximants to the fourth-order series 
for the staggered susceptibility.\protect{\cite{previous}}
The solid circles represent the QMC simulation results.
\protect{\cite{KatoImada,Nonomura}}}
\label{fig:phase_c}%-----------------------------------------
\end{figure}

%%%%%%%%%% dimer-chain %%%%%%%%%%%%
The above analysis may not be  sufficient to discuss the 2D
ladder-structure systems generically,
because the parameter region treated by the cluster expansion is 
restricted. To make a  complementary  analysis,
 we next regard the present system as the coupled-dimer chains, 
for which  the bold, the thin and the dashed lines 
in Fig. \ref{fig:model_d} represent the coupling 
constants $1$, $\alpha_{\rm D}\lambda_{\rm D}$ and 
$\lambda_{\rm D}$.
We carry out the similar calculation up to the eighth order in 
$\alpha_{\rm D}$, and list
the resulting power series for several values $\alpha_{\rm D}$
in Table \ref{tbl:II}.
%%%%%%%%%%%%%%%%%%%%%%%%%%%%%%%%%%%%
\begin{table}
\caption{Series coefficients for the dimer expansion of 
the spin gap $\Delta = E(\pi, 0)$ 
for the coupled-dimer-chain system.}
\begin{tabular}{ccccc}
n & $\alpha_{\rm D}=0.0$ & $\alpha_{\rm D}=0.2$ & 
$\alpha_{\rm D}=0.5$ & $\alpha_{\rm D}=1.0$ \\
\hline
0 & 1.0000000  & 1.0000000  & 1.0000000  & 1.0000000  \\
1 &-0.50000000 &-0.70000000 &-1.0000000  &-1.5000000  \\
2 &-0.37500000 &-0.45500000 &-0.50000000 &-0.37500000 \\
3 & 0.031250000& 0.063250000& 0.062500000& 0.031250000\\
4 &-0.013020833&-0.030220833&-0.067708333&-0.34635417 \\
5 &-0.061930339&-0.10737895 &-0.25710720 &-0.88050673 \\
6 & 0.010735971& 0.040159654& 0.039823179&-0.16209751 \\
7 &0.0030713964&-0.017981720&-0.12801535 &-0.20085681 \\
8 &-0.031547664&-0.077754777&-0.28991249 &-0.63949210 \\
\end{tabular}
\label{tbl:II}
\end{table}
%%%%%%%%%%%%%%%%%%%%%%%%%%%%%%%%%%%%%%
Applying the Pad\'e approximants to the computed spin gap, 
we obtain the phase diagram shown in Fig. \ref{fig:phase_c}.
%%%%%%%%%%%%%%%%%%%%%%%%%%%%%%%%%%%%%%
% -------- the phase diagram for the dimer chain -----------
%%%%%%%%%%%%%%%%%%%%%%%%%%%%%%%%%%%%%
In this figure, 
the solid line represents the phase boundary obtained
by the biased [4/3] Pad\'e approximants,
and the dashed line is the one  obtained previously 
by the staggered susceptibility.\cite{previous}
We find that these two boundaries are in fairly good agreement
with each other,  and furthermore
consistent with those of the QMC simulations\cite{KatoImada,Nonomura} 
(the solid circles in Fig. \ref{fig:phase_c}).  By these 
comparisons, we can say that our cluster expansion approach gives  
quite accurate results for the phase diagram. 
%%%%%%%%%%%%%%%%%%%%%%%%%%%%%%%%%%%%%%%%%%%%%%%%
%%Note that the point $(\alpha_{\rm D}, \lambda_{\rm D})=(0, 1)$ is 
%%just located on the critical line which 
%%separates the magnetically ordered and disordered phases in
%%Fig. \ref{fig:phase_c}. This correctly reproduces the fact that 
%%the ground state of the spin-1/2 Heisenberg chain is 
%%in a critical spin liquid phase
%%without the spin gap.\cite{Des,Affleck}

%%%%%%%%%%%%%%%%%%%%%%%%%%%%%%%%%%%%%%%%%%
\subsection{Plaquette expansion}
%%%%%%%%%%%%%%%%%%%%%%%%%%%%%%%%%%%%%%%%%%%%%%
In the following, we consider the plaquette-structure systems.
Introducing the spin systems with two kinds of the plaquette structures,
we discuss the quantum phase transitions between 
the ordered and the disordered states.
We note here that series expansion studies on plaquette systems 
have been done extensively
by several groups so far,\cite{GelCVO,Fukumoto,metagel,previous} 
which we shall also compare with our results in some special cases.
%%%%%%%%%%%%%%%%%%%%%%
\subsubsection{plaquettes on a square lattice}

\begin{figure}[htb]
\vspace{0.1cm}
\epsfxsize=7cm
\centerline{\epsfbox{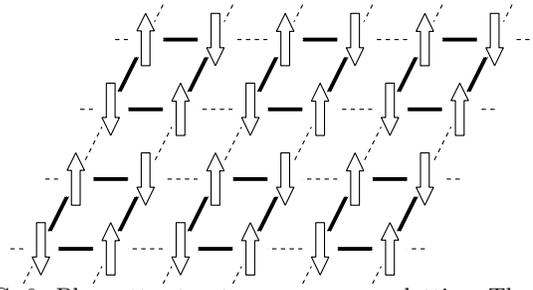}} 
\caption{Plaquette structure on a square lattice.
The bold,  thin, and  dashed lines denote
the coupling constants 1, $\lambda$, and $\alpha\lambda$
among $s=1/2$ spins}
\label{fig:model_p}
\end{figure}

\begin{figure}[htb]
\vspace{0.1cm}
\epsfxsize=7cm
\centerline{\epsfbox{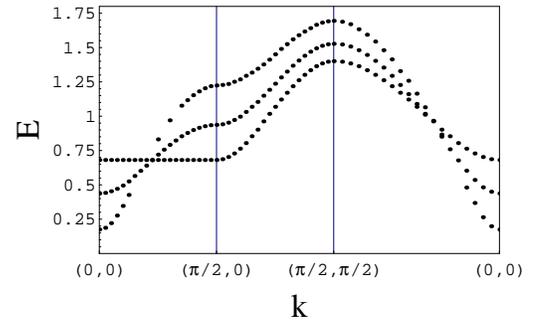}} 
\caption{
Plots of the spin-triplet dispersion $E({\bf k})$ along high-symmetry cuts
through the Brillouin zone for the plaquette structure system 
with the couplings $\alpha=0.0, 0.5, 1.0$ 
(shown in the figure from the top to the bottom at $(0, 0)$,
 respectively) 
in the case $\lambda=0.5$. }
\label{fig:dis_p}%-----------------------------------------
\end{figure}

\begin{figure}[htb]
\vspace{0.1cm}
\epsfxsize=7cm
\centerline{\epsfbox{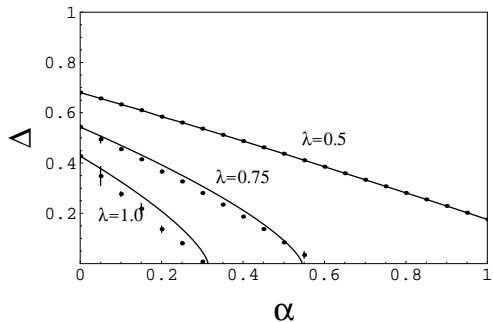}} 
\caption{Spin gap $\Delta=E(0, 0)$ for the plaquette system 
in Fig. \protect{\ref{fig:model_p}}.  The dots denote  
the results computed by
the first order differential method, whereas the solid line  
represents those corrected with  the Pad\'e analysis around 
the critical point.
}
\label{fig:gap_p}%-----------------------------------------
\end{figure}

\begin{figure}[htb]
\vspace{0.1cm}
\epsfxsize=7cm
\centerline{\epsfbox{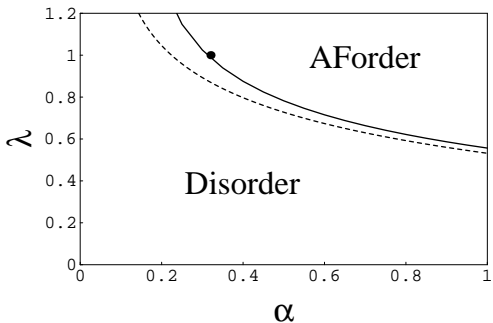}} 
\caption{Phase diagram for the plaquette system on a square lattice.  
The solid line indicates the 
phase boundary obtained by the Dlog [2/1] Pad\'e approximants
and the dashed line is the one obtained previously
by the staggered susceptibility.\protect{\cite{previous}}
The solid  circles denote  the QMC simulation results.
\protect{\cite{ImadaIino}}}
\label{fig:phase_p}%-----------------------------------------
\end{figure}

%%%%%%%%%%%%%%%%%%%%
First, we treat the plaquettes on a square lattice 
shown in Fig. \ref{fig:model_p}.
%%%%%%%%%%%%%%%%%%%%%%%%%%%%%%%%%%%%%%%
% ----------- figure model plaquette on square lattice -------
% ------------------------------------------------
%%%%%%%%%%%%%%%%%%%%%%%%%%%%%%%%%%%%%%
The starting Hamiltonian $H_0$ is composed of the isolated plaquettes,
whose ground state is spin singlet with the excitation gap $\Delta = 1$.
We study how the antiferromagnetic correlation develops
in the presence of the inter-plaquette interaction $\lambda$ and 
$\alpha \lambda$.
In the previous paper,\cite{previous} 
we calculated the staggered susceptibility 
up to the fourth order and  determined the phase boundary between 
the magnetically ordered and the disordered states (see 
the dashed line in Fig. \ref{fig:phase_p}).
We here calculate the dispersion for spin excitations up to the fifth order,
 and list the obtained series of 
the spin gap $\Delta=E(0, 0)$ for several  values of $\alpha$ 
in Table \ref{tbl:III}.
%%%%%%%%%%%%%%%%%%%%%%%%%%%%%%%%%%%%%%%%%%%
\begin{table}
\caption{Series coefficients for the plaquette expansion of 
the spin gap $\Delta = E(0, 0)$ 
for the plaquette system on a square lattice.}
\begin{tabular}{ccccc}
n & $\alpha=0.0$ & $\alpha=0.2$ & 
$\alpha=0.5$ & $\alpha=1.0$ \\
\hline
0 & 1.0000000  & 1.0000000  & 1.0000000  & 1.0000000 \\
1 &-0.66666667 &-0.80000000 &-1.0000000  &-1.3333333 \\
2 & 0.019675926&-0.068425926&-0.19762731 &-0.40509259\\
3 & 0.064935378& 0.016056713&-0.081316913&-0.28178048\\
4 & 0.043061549& 0.019385674&-0.038615359&-0.20391542\\
5 & 0.039539538& 0.030819425&-0.019630017&-0.23535878\\
\end{tabular}
\label{tbl:III}
\end{table}
%%%%%%%%%%%%%%%%%%%%%%%%%%%%%%%%%%%%%%%%%%%%%%%
Using the first-order inhomogeneous differential methods,
we obtain the dispersion relation shown  in Fig. \ref{fig:dis_p}.
%%%%%%%%%%%%%%%%%%%%%%%%%%%%%%%%%
% --------- dispersion relation for plaquette expansion ----------------
%%%%%%%%%%%%%%%%%%%%%%%%%%%%%%%%%
Note that the Brillouin zone is reduced to a quarter of the original one 
due to the plaquette structure.
We also show the spin gap $\Delta=E(0, 0)$ as a function of $\alpha$ 
in Fig. \ref{fig:gap_p}.
%%%%%%%%%%%%%%%%%%%%%%%%%%%%%%%%%
% --------- gap for plaquette expansion ----------------
%%%%%%%%%%%%%%%%%%%%%%%%%%%%%%%%%
As the  inter-plaquette coupling $\alpha$ is increased,
the antiferromagnetic correlation grows up, which causes 
the decrease in  the excitation gap,
and finally induces the quantum phase transition to the antiferromagnetic  
 with the vanishing spin gap.
In the case $(\alpha, \lambda)=(0,1)$,
our model is reduced  to the independent isotropic two-leg ladders for which 
the spin gap $\Delta=0.43$ is obtained.
This value is slightly small compared with
$\Delta=0.504$ (density matrix renormalization group)\cite{DMRG} 
and 0.5028 (dimer expansion),\cite{iso_ladder} which 
implies that  higher-order cluster expansions may be necessary to
obtain more accurate values of the spin gap for 
the plaquette system.  In contrast, it 
is shown below that the phase diagram can be obtained with
much higher accuracy. By
 applying Pad\'e approximants to the power series of the spin gap,
we obtain the phase diagram in Fig. \ref{fig:phase_p}.
%%%%%%%%%%%%%%%%%%%%%%%%%%%%%%%%%%%%%%
% -------- the phase diagram for the plaquette system -----------
%%%%%%%%%%%%%%%%%%%%%%%%%%%%%%%%%%%%%
It is remarkable that the phase boundary 
given in this paper  considerably improves 
the previous one in the small $\alpha$ regime, which can be 
confirmed by the result of the QMC simulations
(the dot shown in the figure).\cite{ImadaIino}
We note  here that for the special case of $\alpha=1$,
similar results were previously 
 reported by Fukumoto {\it et al.}\cite{Fukumoto} 
and Weihong {\it et al.}\cite{metagel}

%%%%%%%%%%%%%%%%%%%%%%%%%%%%%%%%%%%%%%%%%%%%%%%%%%%%%%%%%%%%%%
\subsubsection{plaquettes on a 1/5 depleted square lattice}
%%%%%%%%%%%%%%%%%%%%%%%%%%%%%%%%%%%%%%%%%%%%%%%%%%%%%%%%%%%%%%%

%%%%%%%%%%%%%%%%%%%%%%%%%%%%%%
% -------- model of plaquette chains -------------------------
\begin{figure}[htb]
\vspace{0.1cm}
\epsfxsize=7cm
\centerline{\epsfbox{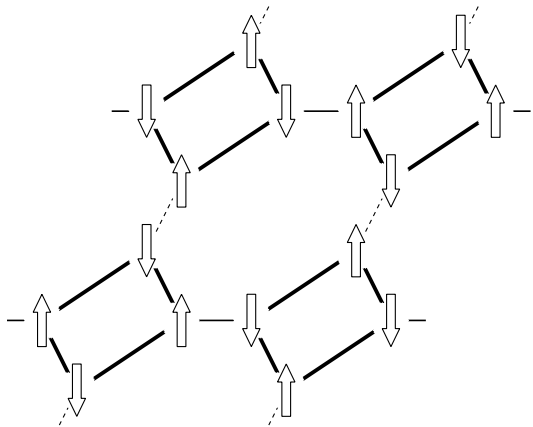}} 
\caption{2D spin system on a 1/5 depleted square lattice.
The bold, the thin, and the dashed lines indicate the coupling constants 
1, $\lambda$, and $\alpha\lambda$ among $s=1/2$ spins.
For simplicity, 
we define the lattice constant as the distance between plaquettes.
 }
\label{fig:model_pc}%-----------------------------------------
\end{figure}
\begin{figure}[htb]
\vspace{0.1cm}
\epsfxsize=7cm
\centerline{\epsfbox{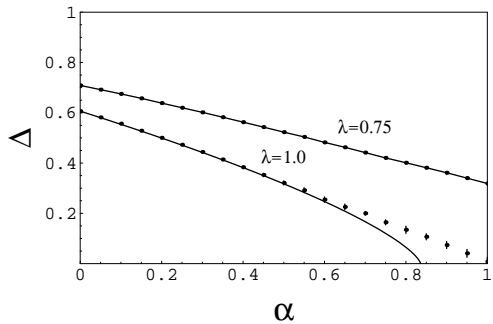}} 
\caption{Spin gap $\Delta=E(\pi, \pi)$ for the plaquette system 
in Fig. \protect{\ref{fig:model_p}}. The dots are the data computed by
the first order differential method, and the solid line 
denotes the corrected one by combining the Pad\'e analysis around 
the region with the small spin gap.
}
\label{fig:gap_pc}%-----------------------------------------
\end{figure}
\begin{figure}[htb]
\vspace{0.1cm}
\epsfxsize=7cm
\centerline{\epsfbox{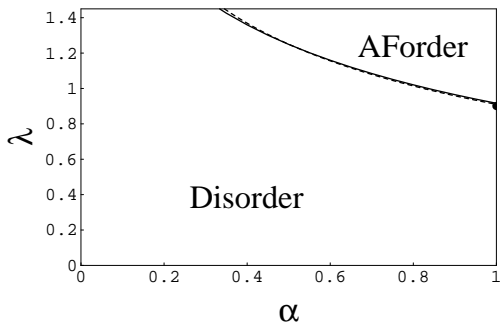}} 
\caption{Phase diagram for the plaquette system  
in Fig. \protect{\ref{fig:model_pc}}.  
The solid line denotes the 
phase boundary obtained by the biased [2/3] Pad\'e approximants
while the dashed line is the previous results\protect{\cite{previous}} 
obtained by the staggered susceptibility.
The solid circles denote the results of the QMC simulation.
\protect{\cite{Troyer}}}
\label{fig:phase_pc2}%-----------------------------------------
\end{figure}

%%%%%%%%%%%%%%%%%%%%%%%%%%%%%%%%%%%%%%%%%
We next deal with the plaquette system shown in Fig. \ref{fig:model_pc},
which may be regarded as 
a 1/5 depleted square lattice,\cite{CaV4O92,Troyer,KatoImadapla}
by extending the work done by Gelfand et al.\cite{GelCVO}
This system is also considered to be  made out of the 
plaquette chains,\cite{KatoImadapla,first,Ivanov,Richter}
since the model with $\alpha=0$ and finite $\lambda$ is reduced to 
the isolated plaquette chains. 
The results for 
the cluster expansion of the spin gap up to the fifth order are tabulated 
in Table \ref{tbl:IV}  for several values of $\alpha$.
%%%%%%%%%%%%%%%%%%%%%%%%%%%%%%%%%%%%%%%%%
\begin{table}
\caption{Series coefficients for the plaquette expansion of 
the spin gap $\Delta = E(\pi, \pi)$ 
for the plaquette system on a 1/5 depleted square lattice. }
\begin{tabular}{ccccc}
n & $\alpha=0.0$ & $\alpha=0.2$ & 
$\alpha=0.5$ & $\alpha=1.0$ \\
\hline
0 & 1.0000000   & 1.0000000   & 1.0000000   & 1.0000000  \\
1 &-0.33333333  &-0.40000000  &-0.50000000  &-0.66666667 \\
2 &-0.10590278  &-0.13236111  &-0.18793403  &-0.32291667 \\
3 & 0.032959989 & 0.024330922 & 0.0092901536&-0.0081862461 \\
4 & 0.024698692 & 0.022081937 & 0.020367093 & 0.033841824 \\
5 &-0.0070637727&-0.0088223796&-0.014159237 &-0.044030516 \\
\end{tabular}
\label{tbl:IV}
\end{table}
%%%%%%%%%%%%%%%%%%%%%%%%%%%%%%%%%%%%%%
The resulting value of $\Delta$ 
deduced by the Pad\'e analysis 
is shown in Fig. \ref{fig:gap_pc} as a 
function  of $\alpha$.
%%%%%%%%%%%%%%%%%%%%%%%%%%%%%%%%%
% --------- gap for plaquette chain ----------------
%%%%%%%%%%%%%%%%%%%%%%%%%%%%%%%%%
In the case of $\alpha=0$, the model is reduced to the isolated plaquette 
chains with the spin gap.  In this case,
by applying the differential methods to the power series,
the spin gap is estimated as $\Delta=0.607\pm0.001$ for $\lambda=1$, 
which is in  good agreement with the result of 
the exact diagonalization $\Delta=0.6086$.\cite{KatoImadapla,Richter}
To observe  the phase transition 
when the couplings $\alpha$  between the plaquette chains increased, 
the phase diagram is shown in Fig. \ref{fig:phase_pc2}.
%%%%%%%%%%%%%%%%%%%%%%%%%%%%%%%%%%%%%%
% -------- the phase diagram for the plaquette chain system -----
%%%%%%%%%%%%%%%%%%%%%%%%%%%%%%%%%%%%%
Here,  the phase boundary (solid line) is determined by applying
the biased [2/3] Pad\'e approximants to the spin gap.
We find that this line is quite consistent with the
 previous results\cite{previous}  shown as the dashed line.
The fact that the two lines evaluated for different quantities 
in different orders produce a quite similar behavior confirms
that the obtained boundary is indeed reliable
although our calculation  is restricted to the lower-order expansions.
We also note that the
results already obtained by QMC\cite{Troyer} and 
also by the plaquette expansion\cite{GelCVO} in the case of $\alpha=1$
are in good agreement with the present one.

%%%%%%%%%%%%%%%%%%%%%%%%%%%%%%%%%%%%%%%%%%%%%%%%%%%%%%%
\subsection{Mixed-spin cluster expansion}
%%%%%%%%%%%%%%%%%%%%%%%%%%%%%%%%%%%%%%%%%%%%%%%%%%%%%%%

Let us now turn to another interesting 2D system composed of  
two kind of different spins,
which has attracted much attention recently. In this mixed-spin system, 
the topological nature of spins is important for the system
to generate the spin gap or induce an 
antiferromagnetic long-range order.
In this subsection, we extend the previous calculations\cite{previous}
to those of the excited states, and 
quantitatively discuss the phase transition in 2D mixed-spin systems.
We will clarify that the arrangement of different spins 
affects the nature of the quantum phase transitions
from the spin-gap phase to the antiferromagnetic  phase.
We shall also check that our series expansion approach
correctly captures the spin structure though our calculation
is based on the lower-order perturbations.
 
\begin{figure}[htb]
\vspace{0.1cm}
\epsfxsize=7cm
\centerline{\epsfbox{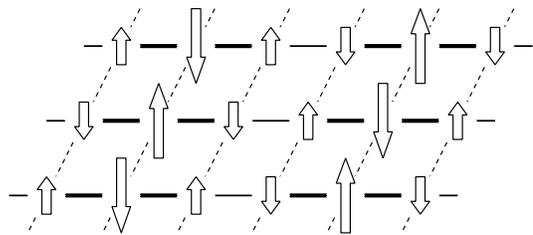}} 
\caption{2D spin system with the {\it columnar}-type  mixed-spin 
structure with $s=1, 1/2.$ We denote
the coupling constants
$1$, $\lambda$, and $\alpha\lambda$.
by  the bold, the thin, and the dashed lines.}
\label{fig:model_col}
\end{figure}

\begin{figure}[htb]
\vspace{0.1cm}
\epsfxsize=7cm
\centerline{\epsfbox{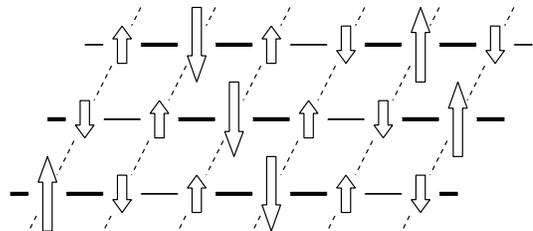}} 
\caption{2D spin system with the {\it diagonal}-type 
mixed-spin structure.
The meanings of 
the bold, the thin, and the dashed lines 
are the same as those in  Fig.
\protect{\ref{fig:model_col}}. 
}
\label{fig:model_dia}
\end{figure}

We deal with two typical systems composed of $s=1/2$ and $1$,
as displayed in 
Figs. \ref{fig:model_col} and \ref{fig:model_dia}. \cite{previous} 
%%%%%%%%%%%%%%%%%%%%%%%%%%%%%%%%%%%%%
% --------- model of alternating spin system 1 ----------
% -------------------------------------------
%%%%%%%%%%%%%%%%%%%%%%%%%%%%%%%%%%%
% --------- model of alternating spin system 2 ------------
% ------------------------------------------------
%%%%%%%%%%%%%%%%%%%%%%%

%%%%%%%%%%%%%%%%%%%%%%%%%%%%%%%%%%%%%%%%%%%%%%%
\subsubsection{columnar-type mixed-spin system}
\begin{figure}[htb]
\vspace{0.1cm}
\epsfxsize=7cm
\centerline{\epsfbox{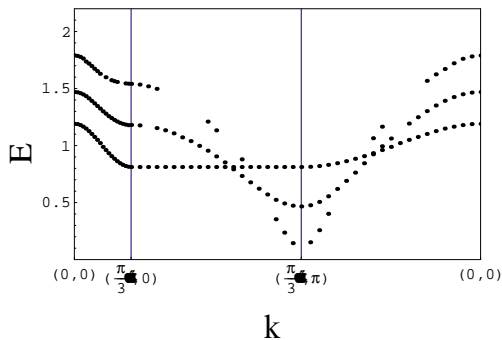}} 
\caption{
Plots of the spin-triplet dispersion relation $E({\bf k})$ 
along high-symmetry cuts through the Brillouin zone for 
the {\it columnar} mixed-spin system with the couplings
$\alpha=0.0, 0.3, 0.6$ 
(shown in the figure from the top to the bottom at $(\pi/3, \pi)$, 
respectively) in the case $\lambda=0.5$.
}
\label{fig:dis_col}%-----------------------------------------
\end{figure}

%%%%%%%%%%%%%%%%%%%%%%%%%%%%%%%%%%%%%%%%%%%%%%
We begin with the columnar-type mixed-spin system, 
for which the mixed-spin chains are stacked  
uniformly in a vertical direction (Fig. \ref{fig:model_col}).
Starting from the mixed-spin clusters of 
$1/2 \circ 1 \circ 1/2$, we perform the series expansion
with respect to $\lambda$.
Note that the Brillouin zone is reduced to a third of the 
original one since the mixed-spin cluster is composed of three spins 
in the $x$-direction. We list the power series obtained 
up to the fifth order for the excitation spectrum in Table \ref{tbl:V}.
It is noted  that in the case of the isolated 
mixed-spin chain ($\alpha=0$),
these coefficients are the same as those for the plaquette chain,
(see Fig. \ref{fig:model_pc})
and thus the isotropic mixed-spin chain with $\lambda=1$ has
the same spin gap $\Delta=0.607$. We can indeed prove that
the mixed-spin chain is identical to the plaquette chain 
as far as the ground state and the low-energy elementary
excitation are concerned. 
%%%%%%%%%%%%%%%%%%%%%%%%%%%%%%%%%
\begin{table}
\caption{Series coefficients for the mixed-spin cluster expansion of 
the spin gap $\Delta = E(\pi/3, \pi)$ 
for the 2D {\it columnar} mixed-spin system.}
\begin{tabular}{ccccc}
n & $\alpha=0.0$ & $\alpha=0.2$ & 
$\alpha=0.5$ & $\alpha=1.0$ \\
\hline
0 & 1.0000000   & 1.0000000   & 1.0000000   & 1.0000000  \\
1 &-0.33333333  &-0.73333333  &-1.3333333   &-2.3333333  \\
2 &-0.10590278  &-0.20868056  &-0.24826389  &-0.008680556\\
3 & 0.032959989 &-0.014472881 &-0.090593252 & 0.50585737 \\
4 & 0.024698692 & 0.013380005 &-0.14493772  &-1.8264592  \\
5 &-0.0070637727&-0.046571788 &-0.44465163  &-5.2417951  \\
\end{tabular}
\label{tbl:V}
\end{table}
%%%%%%%%%%%%%%%%%%%%%%%%%%%%%%%%%
Using the first-order inhomogeneous differential methods,
we obtain the dispersion relation shown in Fig. \ref{fig:dis_col}.
%%%%%%%%%%%%%%%%%%%%%%%%%%%%%%%%%
% --------- dispersion relation for columnar system ----------------
%%%%%%%%%%%%%%%%%%%%%%%%%%%%%%%%%
We recall that the mixed-spin system in the case of $\alpha=0$ 
is reduced to the mixed-spin chain with the spin gap
defined at the wave number ${\bf k}= (\pi/3, \pi)$.
Increasing the inter-chain coupling  $\alpha$,
we can see that the spin gap decreases
as the magnetic correlation grows up, and
finally vanishes at which
%diminishes at which
the phase transition to the magnetically ordered phase takes place.
%%%%%%%%%%%%%%%%%%%%%%%%%%%%%%%%%%%%%%
% -------- the phase diagram  -----------
\begin{figure}[htb]
\vspace{0.1cm}
\epsfxsize=7cm
\centerline{\epsfbox{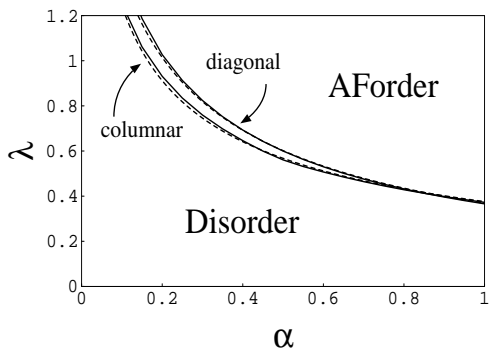}} 
\caption{
The left (right) solid line represents the phase boundary 
for the mixed-spin system in Fig. \protect{\ref{fig:model_col}} (
Fig. \protect{\ref{fig:model_dia}}). 
The results obtained  from the ground-state susceptibility
\protect{\cite{previous}} are shown as the dashed lines.
}
\label{fig:mix}%-----------------------------------------
\end{figure}
%%%%%%%%%%%%%%%%%%%%%%%%%%%%%%%%%%%%%
In Fig. \ref{fig:mix},  the phase boundary is shown by the left solid line,
which is obtained by the biased [2/3] Pad\'e approximants for the spin gap.
In comparison, we  
also display the previous result \cite{previous}  by  the left dashed line, 
which was  determined by  the staggered susceptibility 
in the fourth-order expansion.  
As has been the case for the plaquette systems, we can see again that
 the phase boundaries which were determined 
via the different physical quantities are consistent with each other.
This demonstrates that the reliable phase boundary is established by the
present analysis. 
%%For instance, the critical value is 
%%given by $\alpha=0.17$ for $\lambda=1$.

%%%%%%%%%%%%%%%%%%%%%%%%%%%%%%%%%%%%%%%%%%%%%%%
\subsubsection{diagonal-type mixed-spin system}
\begin{figure}[htb]
\vspace{0.1cm}
\epsfxsize=7cm
\centerline{\epsfbox{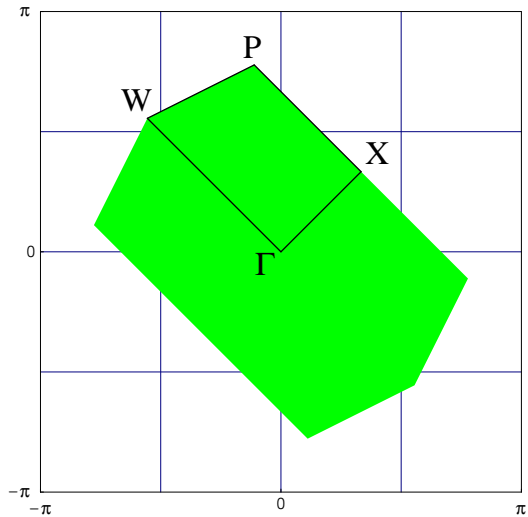}} 
\caption{First Brillouin zone for the 2D system with the {\it diagonal} 
mixed-spin structure.
}
\label{fig:Bril}%-----------------------------------------
\end{figure}

\begin{figure}[htb]
\vspace{0.1cm}
\epsfxsize=7cm
\centerline{\epsfbox{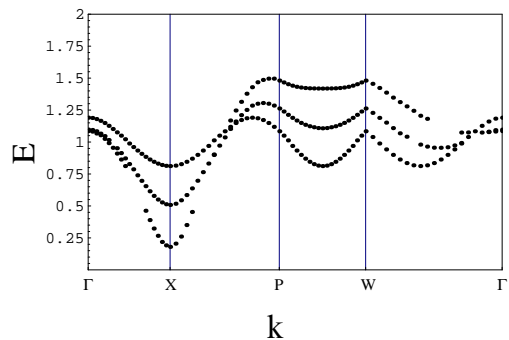}} 
\caption{
Plot of the spin-triplet dispersion relation $E({\bf k})$ 
for the {\it diagonal} mixed-spin system with the coupling parameters
$\alpha=0.0, 0.3, 0.6$ (shown in figure from the top to the bottom
at $(\pi/3, \pi/3)$, respectively) when $\lambda=0.5$.
}
\label{fig:dis_dia}%-----------------------------------------
\end{figure}

%%%%%%%%%%%%%%%%%%%%%%%%%%%%%%%%%%%%%%%%%%%%%%
We next discuss the diagonal-type mixed-spin system
shown in Fig. \ref{fig:model_dia},
for which  the mixed-spin chains are stacked diagonally.
According to this structure, the shape of the Brillouin zone for the 
diagonal system is quite different from 
those for the columnar one as shown in Fig. \ref{fig:Bril}.
%%%%%%%%%%%%%%%%%%%%%%%%%%%%%%%%%%%%%%
% -------- Brillouin zone -----------
%%%%%%%%%%%%%%%%%%%%%%%%%%%%%%%%%%%%%
The definition of the coupling constants is the same 
as that in Fig. \ref{fig:model_col}.
The mixed-spin cluster expansion for the excited states 
up to the fifth order with the asymptotic analysis
yields the phase diagram and the dispersion relations shown 
in Figs. \ref{fig:mix} and \ref{fig:dis_dia}, respectively.
In Fig. \ref{fig:mix}, 
the right solid line represents the phase boundary determined by 
the biased [3/2] Pad\'e approximants for the spin gap .
%%%%%%%%%%%%%%%%%%%%%%%%%%%%%%%%%
% --------- dispersion relation for diagonal system ----------------
%%%%%%%%%%%%%%%%%%%%%%%%%%%%%%%%%
The resulting series for some particular values of $\alpha$ are 
tabulated in Table \ref{tbl:VI}.
%%%%%%%%%%%%%%%%%%%%%%%%%%%%%%%%
\begin{table}
\caption{Series coefficients for the mixed-spin cluster expansion of 
the spin gap $\Delta = E(\pi/3, \pi/3)$ 
for the 2D {\it diagonal} mixed-spin system.}
\begin{tabular}{ccccc}
n & $\alpha=0.0$ & $\alpha=0.2$ & 
$\alpha=0.5$ & $\alpha=1.0$ \\
\hline
0 & 1.0000000   & 1.0000000   & 1.0000000   & 1.0000000  \\
1 &-0.33333333  &-0.66666667  &-1.1666667   &-2.0000000  \\
2 &-0.10590278  &-0.22430556  &-0.42925347  &-0.84375000 \\
3 & 0.032959989 & 0.025971997 &-0.066280914 &-0.41995001 \\
4 & 0.024698692 & 0.0053494949&-0.11649536  &-0.96473863 \\
5 &-0.0070637727&-0.0082040994& 0.020462458 &-0.0043513404\\
\end{tabular}
\label{tbl:VI}
\end{table}
%%%%%%%%%%%%%%%%%%%%%%%%%%%%
For $\alpha=0$, the system correctly reproduces an assembly of 
independent mixed-spin chains which have the disordered ground state.
Increasing $\alpha$, the correlation among the mixed-spin chains grows up, 
and the quantum phase transition occurs.
Especially, in the case of the mixed-spin chains 
with the isotropic bonds $(\lambda=1)$, 
the phase transition to the ordered state occurs 
at the critical value $\alpha_c=0.21$.
We note that as well as the case of the columnar case,
this line is consistent with 
the phase boundary determined from the 
ground-state susceptibility,\cite{previous}
shown as the  dashed line in Fig. \ref{fig:mix}, which
may ensure that phase diagrams for both of 
the two distinct mixed-spin systems are determined 
in rather high accuracy.

\section{Summary}

We have performed the systematic cluster expansion 
to study the two dimensional quantum spin systems with modulated
lattice as well as spin structures.  By applying the asymptotic 
analysis to the obtained series, we have calculated the 
excitation spectrum and have discussed the quantum phase transitions.
We have thus constructed the  
phase diagram which improves  the previous one obtained via 
the staggered susceptibility.
In particular, we find that the present results for the  
systems with the ladder and plaquette structures 
are in fairly good agreement with the results of the 
QMC simulations. We have further studied the critical phenomena for 
the spin systems with modulated $S=1, 1/2$ structure.
By careful study on two types of slightly different systems with 
mixed-spins, it has been clarified that the topology 
of the spin arrangement plays an important role to 
stabilize the spin-gap phase.

%%%%%%%%%%%%%%%%%%%%%%%%%%%%%%%%%%%%%%%%%%%%%%%%%%%%%%%%%%%%%%
\section*{Acknowledgements}
The work is partly supported by a 
Grant-in-Aid from the Ministry of Education, Science, Sports, 
and Culture. A. K. is supported by the Japan Society 
for the Promotion of Science.
Numerical computation in this work was carried out 
at the Yukawa Institute Computer Facility.

%%%%%%%%%%%%%%%%%%%%%%%%%%%%%%%%%%%%%%%%%%%%%%%%%%%%%%%%%%%%%


\begin{thebibliography}{99}

\bibitem{CaV4O91}
S. Taniguchi, T. Nishikawa, Y. Yasui, Y. Kobayashi, M. Sato, 
T. Nishioka, M. Kontani and K. Sano:
J. Phys. Soc. Jpn. {\bf 64} 2758 (1995).

\bibitem{meta}
K. Kodama, H. Harashina, H. Sasaki, Y. Kobayashi, M. Kasai,
S. Taniguchi, Y. Yasui, M. Sato, K. Kakurai, T. Mori and M. Nishi:
J. Phys. Soc. Jpn. {\bf 66} 793 (1997).

\bibitem{CaV4O92}
K. Ueda, H. Kontani, M. Sigrist and P. A. Lee:
Phys. Rev. Lett. {\bf 76} 1932 (1996).

\bibitem{Troyer}
M. Troyer, H. Kontani and K. Ueda:
Phys. Rev. Lett. {\bf 76} 3822 (1996).

\bibitem{KatoImadapla}
N. Katoh and M. Imada: J. Phys. Soc. Jpn. {\bf 64} 4105 (1995).

\bibitem{GelCVO}
M. P. Gelfand, Z. Weihong, R. R. P. Singh, J. Oitmaa and C. J. Hamer:
Phys. Rev. Lett. {\bf 77} 2794 (1996). 

\bibitem{CaV4O93}
O. A. Starykh, M .E. Zhitomirsky, D. I. Khomskii,
R. R. P. Singh and K. Ueda:
Phys. Rev. Lett. {\bf 77} 2558 (1996).

\bibitem{Fukumoto}
Y. Fukumoto and A. Oguchi:
J. Phys. Soc. Jpn. {\bf 67} 697 (1998); 
J. Phys. Soc. Jpn. {\bf 67} 2205 (1998).

\bibitem{metagel}
Z. Weihong, J. Oitmaa and C. J. Hamer:
Phys. Rev. B {\bf 58} 14147 (1998);
R. R. P. Singh, Z. Weihong, C. J. Hamer and J. Oitmaa:
cond-mat / 9904064.

\bibitem{Kageyama}
H. Kageyama, K. Yoshimura, R. Stern, N. V. Mushnikov, K. Onizuka, M. Kato,
K. Kosuge,  C. P. Slichter, T. Goto and Y. Ueda:
Phys. Rev. Lett. {\bf 82} 3168 (1999).

\bibitem{Miyahara}
S. Miyahara and K. Ueda: 
Phys. Rev. Lett. {\bf 82} 3701 (1999).

\bibitem{BO3}
Z. Weihong, C. J. Hamer and J. Oitmaa:
cond-mat/9811030.

\bibitem{Yee}
G. T. Yee, J. M. Manriquez, D. A Dixon, R. S. McLean, D. M. Groski, 
R. B. Flippen, K. S. Narayan, A. J. Epstein and J. S. Miller:
Adv. Mater. {\bf 3} 309 (1991);
Inorg. Chem. {\bf 22} 2624 (1983);
Inorg. Chem. {\bf 26} 138 (1987).

\bibitem{Pati}
S. K. Pati, S. Ramasesha and D. Sen: 
Phys. Rev. B {\bf 55} 8894 (1997);
A. K. Kolezhuk, H.-J. Mikeska and S. Yamamoto: 
Phys. Rev. B {\bf 55} R3336 (1997);
F. C. Alcaraz and A. L. Malvezzi:
J. Phys. A {\bf 30} 767 (1997);
H. Niggemann, G. Uimin and J. Zittartz:
J. Phys. Cond. Matt.: {\bf 9} 9031 (1997).
	
\bibitem{Vega}
H. J. de Vega and F. Woynarovich:
J. Phys. A {\bf 25} 449 (1992);
M. Fujii, S. Fujimoto and N. Kawakami:
J. Phys. Soc. Jpn. {\bf 65} 2381 (1996).

\bibitem{Tonegawa}
T. Tonegawa, T. Hikihara, M. Kaburagi, T. Nishino,
S. Miyashita and H.-J. Mikeska: J. Phys. Soc. Jpn. {\bf 67} 1000 (1998).

\bibitem{Fukui}
T. Fukui and N. Kawakami:
Phys. Rev. B {\bf 55} R14709 (1997);
Phys. Rev. B {\bf 56} 8799 (1997).

\bibitem{first}
A. Koga, S. Kumada, N. Kawakami and T. Fukui:
J. Phys. Soc. Jpn. {\bf 67} 622 (1998).

%%\bibitem{second}
%%A. Koga, S. Kumada and N. Kawakami: J. Phys. Soc. Jpn. 
%%{\bf 68} 642 (1999).

\bibitem{previous}
A. Koga, S. Kumada and N. Kawakami: 
J. Phys. Soc. Jpn. {\bf 68} 2373 (1999).

\bibitem{iso_ladder}
J. Oitmaa, R. R. P. Singh and Z. Weihong:
Phys. Rev. B {\bf 54} 1009 (1996),  % ladder
Z. Weihong, V. Kotov and J. Oitmaa:
Phys. Rev. B {\bf 57} 11439 (1998). % ladder

\bibitem{singh}
R. R. P. Singh, M. P. Gelfand and D. A. Huse:
Phys. Rev. Lett. {\bf 61} 2484 (1988). % first

\bibitem{studychi}
M. P. Gelfand, R. R. P. Singh and D. A. Huse:
J. Stat. Phys. {\bf 59} 1093 (1990). % for study 

\bibitem{studygap}
M. P. Gelfand:
Solid State Commun. {\bf 98} 11 (1996). % for dispersion study

\bibitem{Pade}
A. J. Guttmann, in {\it Phase Transitions and Critical Phenomena},
edited by C. Domb and J. L. Lebowitz 
(Academic, New York, 1989), Vol. 13.


\bibitem{2D}
M. P. Gelfand, R. R. P. Singh and D. A. Huse:
Phys. Rev. B {\bf 40} 10801 (1989); % 2D J1-J2-J3 model
M. P. Gelfand:
Phys. Rev. B {\bf 42} 8206 (1990);  % 2D checkerboard, striped, columnar ..
I. Affleck, M. P. Gelfand and R. R. P. Singh:
J. Phys. A {\bf 27} 7313 (1994).   % chain -> 2D

\bibitem{bilayer}
K. Hida:
J. Phys. Soc. Jpn. {\bf 61} 1013 (1992);
M. P. Gelfand:
Phys. Rev. B {\bf 53} 11309 (1996);% bilayer spectrum
Y. Matsushita, M. P. Gelfand and C. Ishii:
J. Phys. Soc. Jpn. {\bf 66} 3648 (1997); % bilayer with general couplings
Z. Weihong:
Phys. Rev. B {\bf 55} 12267 (1997). % bilayer  various expansion etc.


\bibitem{CHN}
S. Chakravarty, B. I. Halperin and D. R. Nelson:
Phys. Rev. B {\bf 39} 2344 (1989).

\bibitem{Ferer}
M. Ferer and A. Hamid-Aidinejad:
Phys. Rev. B {\bf 34} 6481 (1986).

\bibitem{Rice}
T. M. Rice, S. Gopalan and M. Sigrist:
Europhys. Lett. {\bf 23} 445 (1993);
E. Dagotto and T. M. Rice: 
Science {\bf 271} 618 (1996).

\bibitem{KatoImada}
N. Katoh and M. Imada:
J. Phys. Soc. Jpn. {\bf 63} 4529 (1994).

\bibitem{ImadaIino}
M. Imada and Y. Iino
J. Phys. Soc. Jpn. {\bf 66} 568 (1997).

\bibitem{Nonomura}
Y. Nonomura also studied this system by means of QMC simulations,
(unpublished)

\bibitem{DMRG}
S. R. White, R. M. Noack and D. J. Scalapino:
Phys. Rev. Lett. {\bf 73} 886 (1994).

\bibitem{Ivanov}
N. B. Ivanov and J. Richter:
Phys. Lett. {\bf 232A} 308 (1997).

\bibitem{Richter}
J. Richter, N. B. Ivanov and J. Schulenburg:
J. Phys. Condence Matt. {\bf 10} 3635 (1998).

\end{thebibliography}
\end{document}